\documentclass[11pt,twoside]{article}
\usepackage{asp2010}
\resetcounters

\begin{document}

\title{The Slope of the Upper End of the IMF and the Upper Mass Limit: An Observer's Perspective}
\author{Philip Massey$^1$
\affil{$^1$Lowell Observatory, 1400 W Mars Hill Road, Flagstaff, AZ 86001}}

\begin{abstract}
There are various ways of measuring the slope of the upper end of the IMF.  Arguably the most direct of
these is to place stars on the H-R diagram and compare their positions with stellar evolutionary models.
Even so, the masses one infers from this depend upon the exact methodology used.  I briefly discuss
some of the caveats and go through a brief error analysis.  I conclude that the current data suggest that the IMF slopes are the same to
within the errors.  Similarly the determination of the upper mass ``limit" is dependent upon how well one
can determine the masses of the most massive stars within a cluster.  The recent finding by \citet{CrowtherR136}
invalidates the claim that there is a 150$M_\odot$ upper limit to the IMF, but this is really not surprising given the weakness of the previous evidence.
\end{abstract}

\section{Introduction}

The goal of this talk is to summarize what I think we know (and don't know!) about the
upper end of the IMF and the upper mass limit, with an emphasis on how we know what
we know.   The latter allows us to understand how well we know what we think we know.
When the organizer's invited me to this conference I got hooked  by their statement
that we would ``critically re-evaluate the ensemble of accumulated evidence constraining
the upper end of the IMF."  I come here as observer wishing to remind my colleagues of
the uncertainties and caveats that are sometimes overlooked.  Or to put it in NPR's Car Talk
parlance, I'm here to administer an observer's dope slap, where a dope slap is ``an
attention getting device."

\section{IMF Slopes of Clusters Containing Massive Stars}
Let me begin with the most ``direct" way of measuring the IMF: counting stars in the
Hertzsprung-Russell diagram (HRD).

Table~\ref{tab:gammas} lists some IMF slope determinations from \citet{MasseyIMF98}, updated
with more recent values for R136 \citep{MasseyHunter98} and h and $\chi$ Persei \citep{Slesnick}. Included are the ages inferred for the clusters and the masses of the highest mass present. 
These numbers were all determined in a uniform way, and there are certainly differences
between the IMF slopes $\Gamma$'s that are larger than the quoted errors.  But seriously,
where do these numbers come from, and how much can one believe them?  Do these
numbers demonstrate that the IMF slope indeed varies at the upper end?

\begin{table}[!ht]
\caption{\label{tab:gammas} Ages, Upper Mass, and IMF Slopes}
\smallskip
\begin{center}
{\small
\begin{tabular}{lcrcl}
\tableline
\noalign{\smallskip}
Region & Age & $M_{up}$ & $\Gamma$ & Ref.\ \\
&[Myr] & [$M_\odot$] \\
\noalign{\smallskip}
\tableline
\noalign{\smallskip}
\multicolumn{5}{c}{SMC}\\
NGC 346 & 2-4 & 70 & $-1.3\pm0.1$ & \citet{MasseyN346}\\
\multicolumn{5}{c}{LMC}\\
LH 6 & ...  & 85 & $-1.7\pm0.4$& \citet{Oey96}\\
LH 9 & 1-5 & 55 & $-1.4\pm0.2$& \citet{ParkerLH9}\\
LH 10 & 0-3 & 90 & $-1.1\pm0.1$&\citet{ParkerLH9}\\
LH 38 & ... & 85: & $-1.6\pm0.2$ & \citet{Oey96}\\
LH 47/48 & 2-3 & 50 & $-1.3\pm0.2$& \citet{OeyLH47}\\
LH 58 & 2-4 & 50 & $-1.4\pm0.2$&\citet{GarmanyLH58}\\
LH 73 & ... & 65: & $-1.3\pm0.4$&\citet{Oey96}\\
LH 83 & ... & 50: & $-1.3\pm0.5$&\citet{Oey96}\\
LH 114 & ... & 50: & $-1.0\pm0.1$&\citet{Oey96}\\
LH 117/118 & 1-3 & 100 & $-1.6\pm0.2$&\citet{MasseyLH117}\\
R136            & 1-2 & $>$120 & $-1.3\pm0.2$&\citet{MasseyHunter98}\\
\multicolumn{5}{c}{Milky Way}\\
NGC 6823 & 2-7 & 40 & $-1.3\pm0.4$&\citet{MasseyNorth}\\
NGC 6871 & 2-5 & 40 & $-0.9\pm0.4$&\citet{MasseyNorth}\\
NGC 6913 & 4-6 & 40 & $-1.1\pm0.6$&\citet{MasseyNorth}\\
Berkeley 86 & 2-3 & 40 & $-1.7\pm0.4$&\citet{MasseyNorth}\\
NGC 7235 & (6-11) & 15 & ($-$2.0)&\citet{MasseyNorth} \\
NGC 7380 & 2 & 65 & $-1.7\pm0.3$&\citet{MasseyNorth}\\
Cep OB5 & 2-4 & 30 & $-2.1\pm0.6$&\citet{MasseyNorth}\\
IC 1805 & 1-3 & 100 & $-1.3\pm0.2$&\citet{MasseyNorth}\\
NGC 1893 & 2-3 & 65 & $-1.6\pm0.3$&\citet{MasseyNorth}\\
NGC 2244 & 1-3 & 70 & $-0.8\pm0.3$&\citet{MasseyNorth}\\
NGC 6611 & 1-5 & 75 & $-0.7\pm0.2$&\citet{Tuesday}\\
CygOB2 & 1-4 & 110 & $-0.9\pm0.2$&\citet{MasseyCygOB2}\\
Tr 14/16 & 0-3 & $>$120 & $-1.0\pm0.2$ & \citet{MasseyTr14}\\
h and $\chi$ Per & 12.8 & 35 & $-1.3\pm0.2$&\citet{Slesnick}\\
\noalign{\smallskip}
\tableline
\end{tabular}
}
\end{center}
\end{table}

The errors quoted on the IMF slopes $\Gamma$ in the above table are strictly the {\it fitting} uncertainties to the
number of stars as a function of mass.  They do not take into account any of the other errors, such as the determination of
the masses, or for that matter, the degree of coevality that is needed to equate the slope of the present day mass function
with that of the initial mass function.  Let us consider at some length just one of these issues, namely the determination of the
masses.
\subsection{Determining the Masses}
To count the number of stars as a function of mass, one needs to somehow get the masses.
To do this, we must find the star's bolometric luminosity and then rely upon stellar
evolutionary models to convert the luminosity to mass.  
Generally one begins by obtaining {\it UBV} photometry of the stars.  These are used to
identify the hot, luminous stars, usually by the Johnson reddening-free $Q$ index
$(U-B)-0.72(B-V).$  One then takes spectra of as many stars as practical, and uses
each star's spectral type to convert to effective temperature $T_{\rm eff}$.  The effective temperatures
yield the bolometric correction as a function of spectral subtype. Comparison of a star's
observed colors with that expected based on the spectral type and luminosity class 
also yields the color
excess and hence the correction of $V$ for reddening.  In the cases where the distances
are poorly known (as in Galactic clusters) the spectral types can also be used to determine
a ``spectroscopic parallax" to the region, i.e., using the observed de-reddened magnitude
and the absolute visual magnitude expected from each star's spectral type.  For clusters older
than about 10 Myr it is more accurate to use main-sequence fitting from the photometry
itself to determine the distance modulus \citep[as we did for h and $\chi$ Per; see][]{Slesnick}.

So, are spectra really necessary?  Couldn't we just go from $Q$ to $T_{\rm eff}$ and be done with it?
Let's consider the associated errors, following \citet{MasseyIMF98}.    First, assume that the mass-luminosity
relationship goes something like $L\sim m^3$, in accord with the solar-metalllity Geneva evolution models for stars with masses $>60M_\odot$. (Below this mass the exponent is larger.) 
Then $\Delta \log m \sim 0.3 \Delta \log L$.  If we express the luminosity in terms of the bolometric magnitude $M_{\rm bol}$,
then $\Delta \log m \sim 0.1 \Delta M_{\rm bol}$.

The uncertainty in $M_{\rm bol}$ will be a combination of the error in the reddening, the error in the distance, 
and the error in the bolometric correction BC.
The latter is related to the effective temperature $T_{\rm eff}$ as ${\rm BC}=-6.90\log T_{\rm eff} + 27.99$, so
$\Delta {\rm BC}=-6.9 \Delta  \log T_{\rm eff},$ according to \citet{MasseyOStarsII}.

Let us now see how the uncertainties in $\log m$ stack up.  The typical uncertainty in the distance is 0.1 mag, and so that
is one source of uncertainty on $\log M_{\rm bol}$, but of all of the errors that one will be systematic
for stars of all masses in a cluster and will have the least overall effect on the derived IMF slope,
but it will affect the derived masses so let's include it.
  The reddening might be uncertain by a similar amount, and in this case it can be either
  systematic or vary star-by-star.
{\it If} we are using spectral types to get $\log T_{\rm eff}$, then an uncertainty of 1 spectral type corresponds to an uncertainty of
about 0.01 in $\log T_{\rm eff}$ and hence 0.1 in $\log M_{\rm bol}$.  Thus all three terms contribute similarly, and the total uncertainty
in $M_{\rm bol}$ is about 0.2~mag.  The uncertainty in the mass (assuming the evolutionary tracks are gospel) is then 0.02 dex, which is
really very good!  (We will talk about systematics below.) 

\begin{figure}[!ht]
\plottwo{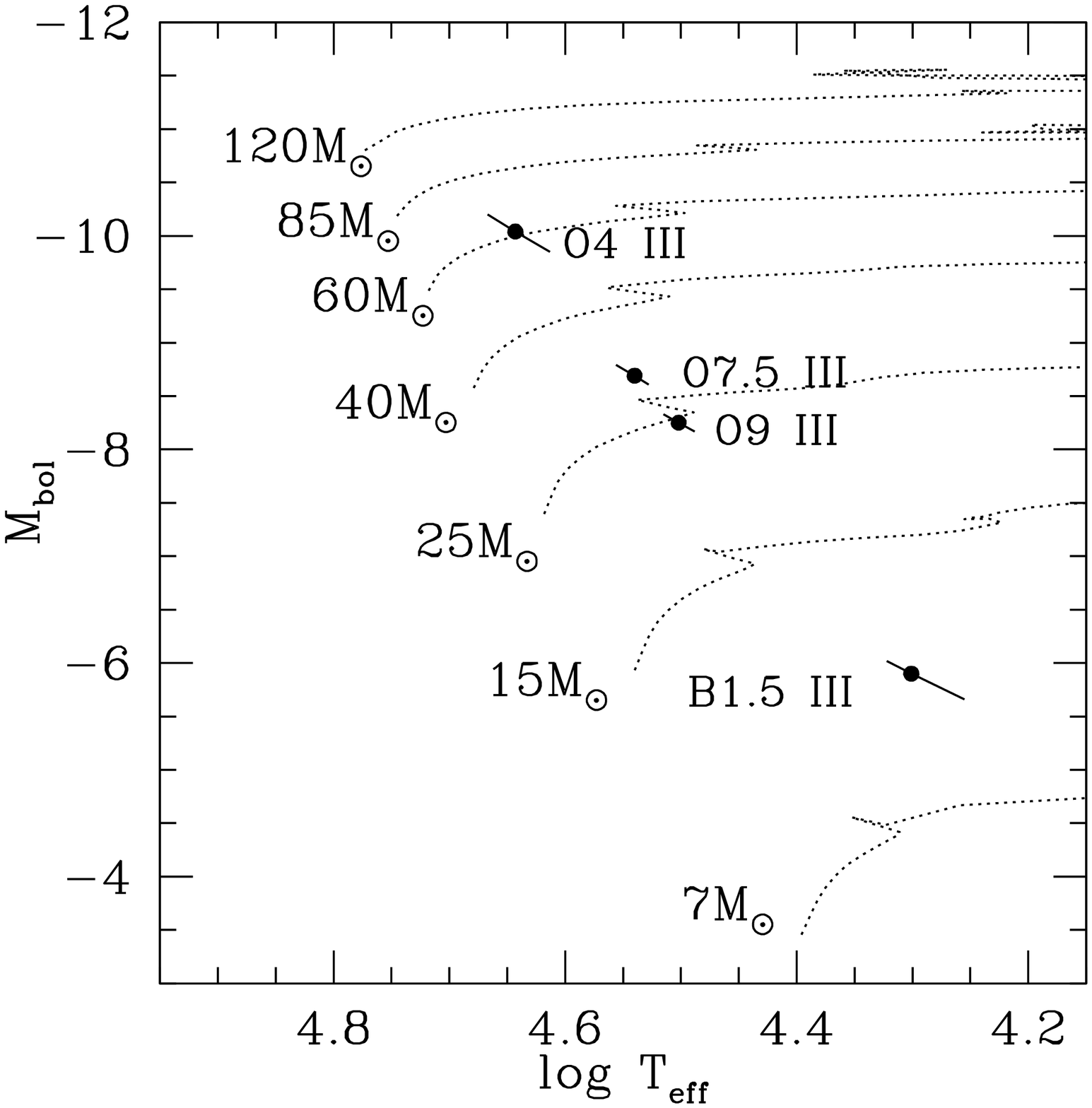}{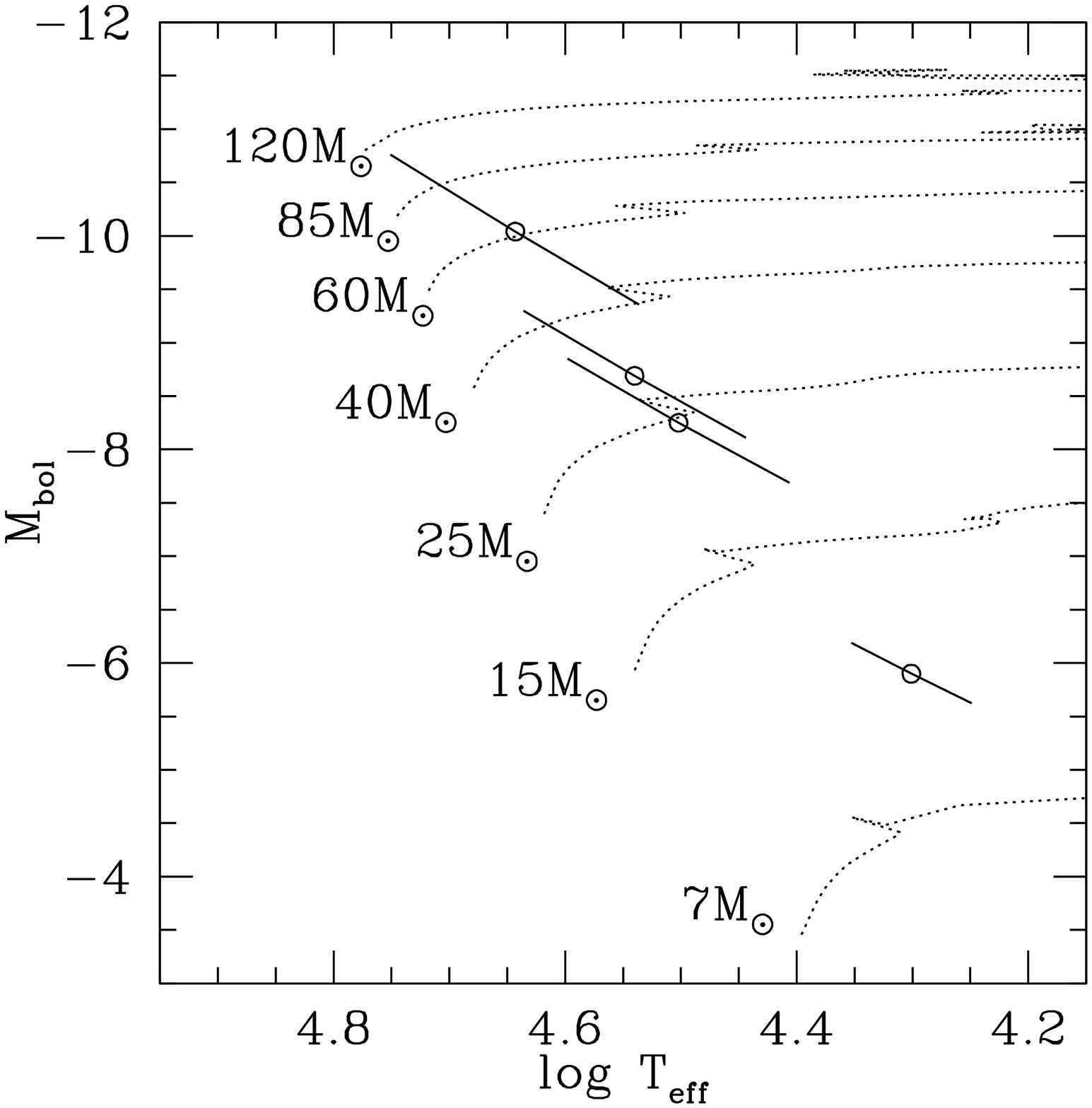}
\caption{\label{fig:cramit} On the left is shown the errors associated with placing stars on the H-R diagram using their spectral types to determine their masses assuming an uncertainty of one spectral subtype.
 On the right is shown the errors associated with placing the same stars based upon their photometry, assuming an uncertainty of 0.05 in $Q$ (i.e., roughly 0.02 mag error each in (U-B) and (B-V).}
\end{figure}

If instead we had depended upon the photometry to derive a mass, we would have to have related a reddening-free index
(such as $Q$) to $\log T_{\rm eff}$ using stellar atmospheres.  For hot, massive stars this is a very frustrating endeavor, as the
effective temperatures of these stars are so high that there is very little difference in the colors of a 50,000 K star and a 40,000 K star,
even though the spectra are quite different.  \citet{MasseyIMF98} derives values for $Q$ and other reddening-free indices, including
ones derived from spacecraft UV-filters.  I found that the standard {\it UBV} $Q$ would vary by only 0.007~mag over the
same temperature range.  In general, for hot stars, $\Delta \log T_{\rm eff} \sim 5 \Delta Q$.   So, a (reasonable) error of 0.05~mag in $Q$
would translate to an uncertainty of 0.25 dex in $\log T_{\rm eff}$, and (via the equation for the BC), about 1.7~mag in $\log M_{\rm bol}$, or
0.2~dex in $\log m$, about 10$\times$ the error from using spectral types.  You can't really tell a 100$M_\odot$ from a 60$M_\odot$ star
using photometry alone. (I'll briefly note that the UV filters fare a bit better but not much.)
This is all better illustrated in Figure~\ref{fig:cramit}, where the relative errors of placing hot, massive stars in the H-R diagram is shown.

How would these errors translate into errors in the IMF slope $\Gamma$?  If the errors were somehow systematic in $\log m$ they would
actually have no effect.  But far more likely there would be biases introduced as a function of mass, as the hottest stars (where the 
dependence upon photometry is even less) are also the most massive.

\begin{figure}[ht]
\plotfiddle{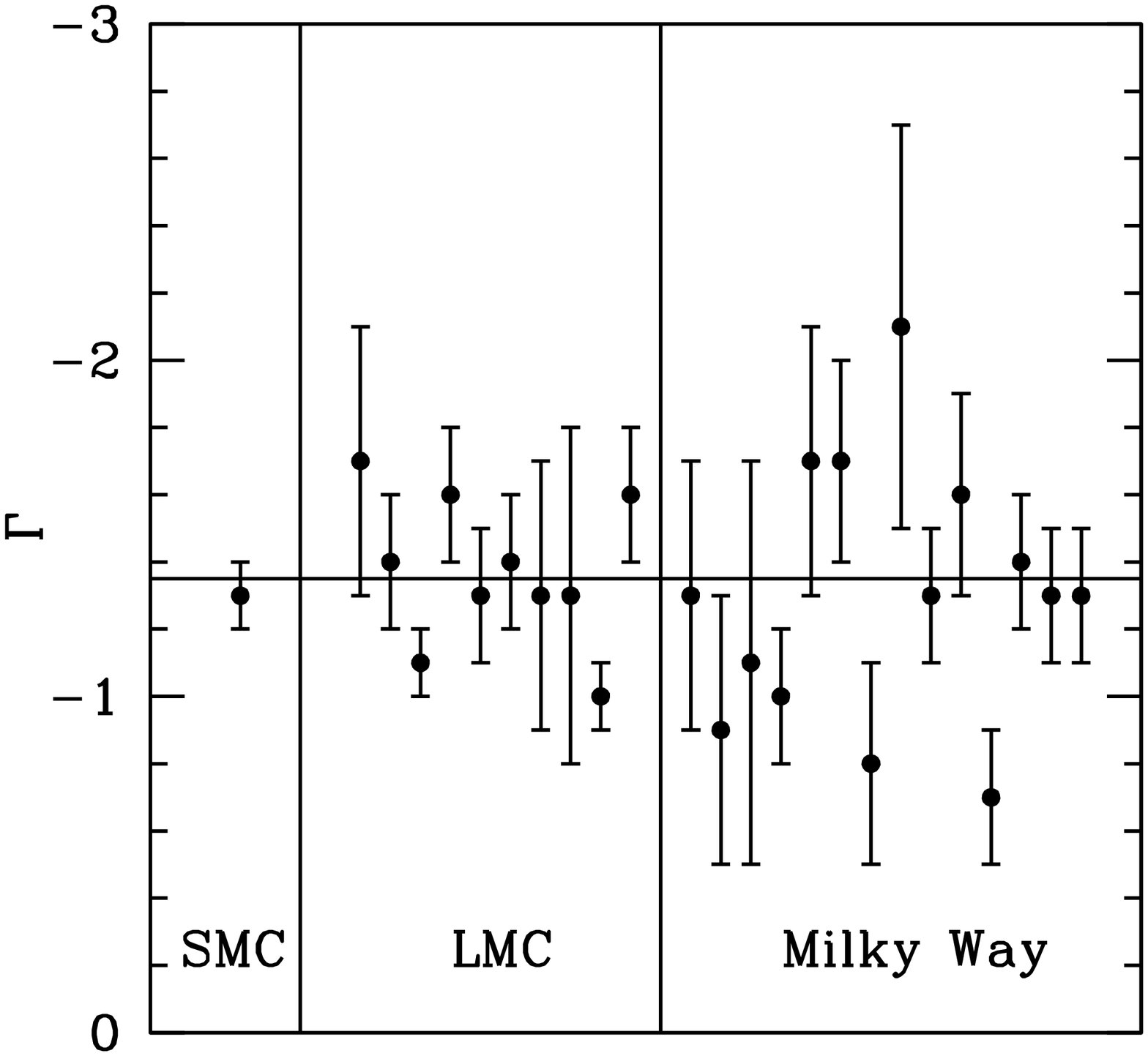}{3.5in}{0}{50}{50}{-150.}{-80.}
\caption{\label{fig:IMFslopes} IMF slopes determined in as consistent a manner as possible
are shown for OB associations and clusters in the SMC, LMC, and Milky Way.  The errors
are undoubtably larger for the Milky Way regions due to the additional reddening and 
uncertainties in distances.  The data come from \citet{MasseyIMF98}, \citet{Slesnick}, and
\citet{MasseyHunter98}, and cover a range of 4 in metallicity and factors of several hundred
in stellar density. A Salpeter IMF slope of $\Gamma=-1.35$ is shown for comparison.}
\end{figure}

We can demonstrate this by just comparing how different researchers using different methods agree on the IMF slope.  Consider
the value for LH 58, a beautiful OB association in the LMC.  It was analyzed by ourselves \citep{GarmanyLH58, MasseyIMF98} using
spectroscopy, and by \citet{Hill94} using their own photometry but no spectroscopy.  \citet{GarmanyLH58} derive $\Gamma=-1.7\pm0.3$,
while \citet{MasseyIMF98} derives $\Gamma=-1.4\pm0.2$ using the identical data, but treating the reddening just a little bit
differently.  Using only photometry, 
\citet{Hill94} find $\Gamma=-2.5\pm0.3$.  \citet{MasseyIMF98} notes that reanalyzing the cluster adopting the \citet{Hill94} conversion
from photometry to effective temperatures, but using the \citet{GarmanyLH58} photometry, leads to $\Gamma=-2.0.$  The conclusion
I take away from this is that reasonable people using reasonable data may still derive IMF slopes that disagree by a significant amount!
If one is going to look for variability of the upper end of the IMF this way one has to analyze the data {\it consistently.}  Figure~\ref{fig:IMFslopes}
shows the IMF slopes from Table~1, where the data have been treated as much the same as possible.  
It's hard to come away from
this thinking there's evidence for a variable IMF slope.

One should recall that  while these values for $\Gamma$ have been found {\it consistently}, that doesn't mean that they are {\it right}
in some absolute sense.  For one thing, binarity has been ignored.  The relative values are still right {\it if} the binary frequency and mass ratio distributions are the same from region to region.  But the actual values shouldn't be.  Furthermore, our conversions have relied upon 
a set of conversions from spectral types to effective temperatures.  That scale is probably uncertain by 10\% in an absolute sense \citep{Conti88,MasseyOStarsII}.

I should also note that the way we went about determining the IMF slope once we had the
masses was a bit crude.  As \citet{MeyerARAA} notes, one could do perhaps better by a
more sophisticated analysis, comparing a chi-square goodness of fit with models of the
cumulative distributions drawn from various IMF slopes. 

\begin{figure}[!ht]
\epsscale{0.5}
\plotone{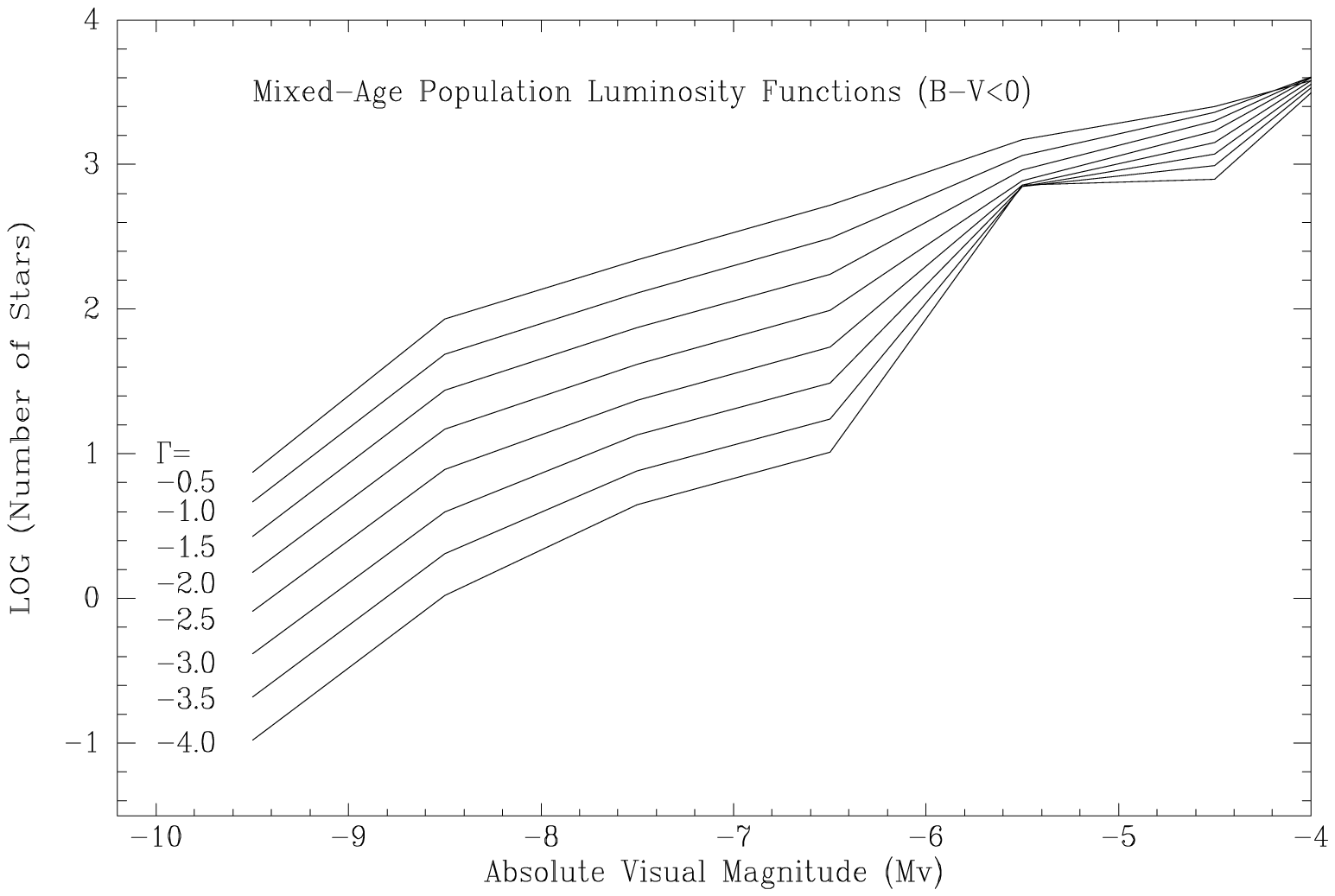}
\caption{\label{fig:lumfunct} The luminosity functions are amazingly insensitive to the IMF slopes, as demonstrated for a mixed-age
population.  From \citet{MasseyIMF98}.}
\end{figure}

What about determining the IMF slope from just a luminosity function?  Well, that turns out to be incredibly insensitive to the value of $\Gamma$,
at least in the case of mixed-age populations.
Consider the case of looking at massive stars in a nearby galaxy.  One can't use all of the stars in a color-magnitude diagram as the vast
majority of these will be dominated by yellow foreground stars \citep[see, for example,][]{MasseyHRD,Drout}.  One can restrict the sample to
the bluest stars only, say, $B-V<0$.  But the sensitivity to the IMF slope is almost non-existent, as shown in Figure~\ref{fig:lumfunct}.  
This works a lot better for stars that are coeval, but without first constructing an H-R diagram it's not clear how one knows {\it a priori} that
a region is coeval. 

\section{The Upper Mass ``Limit"}

\citet{MasseyIMF98} wrote that the highest mass stars identified in OB clusters and associations were consistent with what
one expected by extrapolating the IMF slopes of the lower mass stars: that generally any ``truncation" in the mass function
was simply due to age.   In contrast, \citet{OeyClarke05} argue that by combining data from many regions (mostly those in
Table~1) that one should expect purely by chance to have seen higher mass stars than what have been observed so far, and
argue that stars more massive than 150$M_\odot$ must not be able to form.

Let us think about what the requirements would have to be to find a (say) 300$M_\odot$.  First, my own expectation would
be that we would find it in a rich association, not a sparse one.   Not everyone would agree with that point, as there is another
school of thought that somehow the probability of forming a star of mass $m$ is independent of physical conditions.  I confess I've
never understood that argument, but that doesn't mean it's wrong.
\citet{MeyerARAA} presents nice arguments relating the mass of the highest mass
star seen to  cluster
size. But, regardless of whether we all agree on this point or not, we probably {\it can}
all agree that if were hoping to find 
a very high mass star in a cluster, we would probably do better to look at clusters that were sufficiently young to still have such
a star among the living.

How young is young enough?  Here we are on somewhat shaky grounds, as the best vetted models are those of the Geneva
group \citep[for instance,][]{Meynet05,Meynet03,Charbonnel93,Schaerer93,Schaller92,Maeder89, Maeder88}
and they extend in mass only up to 120$M_\odot$, although Georges Meynet (2010, private
communication) says that there is no reason why higher mass models couldn't be made
available if there was interest in these.
 What is the expected lifetime of a 300$M_\odot$ star?  As one approaches
the Eddington limit, $L~\sim m$ rather than some steeper power \citep{StanTheMan}, and since the main-sequence lifetime $\tau_{\rm ms}$
is going to go something like $\tau_{\rm ms} \sim m/L$ we would expect the lifetimes to be fairly invariant with mass.  At 80-120$M_\odot$ $\tau\sim m^{0.5}.$  The lifetime of a 120$M_\odot$ star is 3.1~Myr at solar metallicity \citep{Meynet03}.  So, one would expect a lifetime somewhere
between around 2~Myr for a 300$M_\odot$ star.

Examination of Table~1 reveals that there are {\it very} few clusters and associations we know of that are that young.  So, the first
thing to consider is that if one is going to perform a statistical test along the lines of \cite{OeyClarke05} then one really has to restrict
one's self only to these few regions that are young enough.

The second point I wish to make is that we really don't know the masses of the highest mass stars in these associations very well.
Take the example of R136a1.  \citet{MasseyHunter98} makes a very conservative estimate of 150$M_\odot$ for its mass, but this
required again extrapolating the mass-luminosity relationship further than the models go.  As they note, the effective temperatures
of such ``Of stars on steroids" (which exhibit hydrogen-rich WN-type emission spectra)
are very poorly known, and without a good effective temperature you don't know the bolometric
luminosity very well, and without the bolometric luminosity you can't get the mass even by extrapolating the mass-luminosity relationship.
Since the time of the conference,  \citet{CrowtherR136} has made a new effective temperature determination by obtaining
VLT optical data and fitting it (along with archival {\it HST} data).  They obtain a much higher temperature (53,000 K) than that found
in the analysis by \citet{HeapR136} (45,000 K), along with a correspondingly higher luminosity and higher mass.  \citet{CrowtherR136} now estimate
the (initial) mass of R136a1 be 300$M_\odot$.  I was tickled when I first saw the paper because during my talk I noted that one could
argue for a higher temperature for this star and that indeed that a modern analysis with CMFGEN might well reveal it to be a 300$M_\odot$
and not 150$M_\odot$.  Bazinga!  So, this is a very well done and interesting study, but when interviewed by the press I got to say
that I wasn't  surprised by it.

Even so, this is hardly the final word on the subject.  The stars in the core of R136 are quite crowded, and \citep{HeapR136} argue that their optical {\it HST} spectrum is contaminated by neighboring stars. It is doubtful that the ground based spectrum,
obtained even with AO, is without contamination. De Koter et al.\ (1997) also present strong arguments
as to why the effective temperatures must be less than 47500 K.  But I'll let the
stellar atmosphere pundits slug that one out.)  How significant this contamination may be, no one knows yet.  Similarly, there is always the possibility
that this star is an undetected binary.  But, even so, it pretty well puts to rest the 150$M_\odot$ 
upper mass limit myth that has existed for the past few years.

The final point to make about the upper mass limit is to remind the reader that age can do a very nice job truncating a luminosity 
function---the truncation does not have to be due to an upper mass cutoff.  The Arches cluster provides an example of this.
\citet{Figer04} cited the truncation of the K-band luminosity function to be proof that there was an upper mass cutoff of about 150$M_\odot$:
extrapolating the IMF suggested there should be 18 stars with masses $>130M_\odot$ while none were found.  But the picture was greatly
clarified thanks to spectroscopy.  \citet{Martins08} obtained spectra of stars, placed them on an HRD and concluded that ``All stars are 2-4 Myr old".
So, one {\it expects} that the luminosity function (and mass function) will be truncated.  The region is just a bit too old to shed any light on
the upper mass limit.

I will say that my prejudice is that there certainly is {\it some} upper mass limit. {\it Something} must limit the mass of a massive star.
Our colleagues in the high mass star formation business here at this conference may tell us what.

\section{Summary}
This conference has certainly been held in a beautiful place!  But I think Sedona also serves as a warning to us: having a number of pieces
of indirect, weak evidence doesn't prove the variability of the upper end of the IMF or that 150$M_\odot$ is the upper mass limit, any more
than anecdotal stories prove the existence of UFOs and cosmic vortexes.  They may be fun to believe in, but one has to look at the
evidence critically.

\acknowledgements  I'm grateful to Deidre Hunter and Michael Meyer for critical comments on the
talk and manuscript, and for many useful discussions over the years.  I'm also grateful to the
conference organizers for bringing us all together here for such a pleasant meeting.

\bibliography{massey_p}

\begin{thebibliography}{}
\expandafter\ifx\csname natexlab\endcsname\relax\def\natexlab#1{#1}\fi
\expandafter\ifx\csname url\endcsname\relax
  \def\url#1{\texttt{#1}}\fi
\expandafter\ifx\csname urlprefix\endcsname\relax\def\urlprefix{URL }\fi
\providecommand{\eprint}[2][]{\url{#2}}

\bibitem[{Bastian et~al.(2010)Bastian, Covey, \& Meyer}]{MeyerARAA}
Bastian, N., Covey, K.~R., \& Meyer, M.~R. 2010, ARAA, 48, in press,
  arXiv:1001.2965v2

\bibitem[{Charbonnel et~al.(1988)Charbonnel, Meynet, Maeder, Schaller, \&
  Schaerer}]{Charbonnel93}
Charbonnel, C., Meynet, G., Maeder, A., Schaller, G., \& Schaerer, D. 1988,
  A\&AS, 101, 415

\bibitem[{Conti(1998)}]{Conti88}
Conti, P. 1998, in O Stars and Wolf-Rayet Stars, edited by P.~S. Conti, \&
  A.~B. Underhill (Washington, D. C.: NASA), vol. 497 of NASA Special
  Publications, 119

\bibitem[{Crowther et~al.(2010)Crowther, Schnurr, Hirschi, Yusof, Parker,
  Goodwin, \& Kassim}]{CrowtherR136}
Crowther, P., Schnurr, O., Hirschi, R., Yusof, N., Parker, R.~J., Goodwin,
  S.~P., \& Kassim, H.~A. 2010, MNRAS, in press, arXiv:1007.3284

\bibitem[{de~Koter et~al.(1997)de~Koter, Heap, \& Hubeny}]{HeapR136}
de~Koter, A., Heap, S.~R., \& Hubeny, I. 1997, ApJ, 477, 792

\bibitem[{Drout et~al.(2009)Drout, Massey, Meynet, Tokarz, \& Caldwell}]{Drout}
Drout, M.~R., Massey, P., Meynet, G., Tokarz, S., \& Caldwell, N. 2009, ApJ,
  703, 441

\bibitem[{Figer(2005)}]{Figer04}
Figer, D.~F. 2005, Nature, 434, 192

\bibitem[{Garmany et~al.(1994)Garmany, Massey, \& Parker}]{GarmanyLH58}
Garmany, C.~D., Massey, P., \& Parker, J.~W. 1994, AJ, 108, 1256

\bibitem[{Hill et~al.(1998)Hill, Madore, \& Freedman}]{Hill94}
Hill, R.~J., Madore, B.~F., \& Freedman, W.~L. 1998, ApJ, 429, 192

\bibitem[{Hillenbrand et~al.(1993)Hillenbrand, Massey, Strom, \&
  Merrill}]{Tuesday}
Hillenbrand, L.~A., Massey, P., Strom, S.~E., \& Merrill, K.~M. 1993, AJ, 106,
  1906

\bibitem[{Maeder \& Meynet(1988)}]{Maeder88}
Maeder, A., \& Meynet, G. 1988, A\&AS, 76, 411

\bibitem[{Maeder \& Meynet(1989)}]{Maeder89}
--- 1989, A\&A, 210, 155

\bibitem[{Martins et~al.(2008)Martins, Hillier, Paumard, Eisenhauer, Ott, \&
  Genzel}]{Martins08}
Martins, F., Hillier, D.~J., Paumard, T., Eisenhauer, F., Ott, T., \& Genzel,
  R. 2008, A\&A, 478, 219

\bibitem[{Massey(1998)}]{MasseyIMF98}
Massey, P. 1998, in The Stellar Initial Mass Function (38th Herstmonceux
  Conference), edited by G.~Gilmore, \& D.~Howell (San Francisco: ASP), vol.
  142 of ASP Conf.\ Ser.\, 17

\bibitem[{Massey(2010)}]{MasseyHRD}
--- 2010, in Hot and Cool: Bridging Gaps in Massive Star Evolution, edited by
  C.~Leitherer, P.~D. Bennett, P.~W. Morris, \& J.~T. Van~Loon (San Francisco:
  ASP), vol. 425 of ASP Conf.\ Ser.\, 3

\bibitem[{Massey \& Hunter(1998)}]{MasseyHunter98}
Massey, P., \& Hunter, D.~A. 1998, ApJ, 493, 180

\bibitem[{Massey \& Johnson(1993)}]{MasseyTr14}
Massey, P., \& Johnson, J.~. 1993, AJ, 105, 980

\bibitem[{Massey et~al.(1995)Massey, Johnson, \&
  DeGioia-Eastwood}]{MasseyNorth}
Massey, P., Johnson, K.~E., \& DeGioia-Eastwood, K. 1995, ApJ, 454, 151

\bibitem[{Massey et~al.(1989{\natexlab{a}})Massey, Parker, \&
  Garmany}]{MasseyN346}
Massey, P., Parker, J.~W., \& Garmany, C.~D. 1989{\natexlab{a}}, AJ, 98, 1305

\bibitem[{Massey et~al.(2005)Massey, Puls, Pauldrach, Bresolin, Kudritzki, \&
  Simon}]{MasseyOStarsII}
Massey, P., Puls, J., Pauldrach, A. W.~A., Bresolin, F., Kudritzki, R.~P., \&
  Simon, T. 2005, ApJ, 627, 477

\bibitem[{Massey et~al.(1989{\natexlab{b}})Massey, Silkey, Garmany, \&
  DeGioia-Eastwood}]{MasseyLH117}
Massey, P., Silkey, M., Garmany, C.~D., \& DeGioia-Eastwood, K.
  1989{\natexlab{b}}, AJ, 97, 107

\bibitem[{Massey \& Thompson(1991)}]{MasseyCygOB2}
Massey, P., \& Thompson, A.~B. 1991, AJ, 101, 1408

\bibitem[{Meynet \& Maeder(2003)}]{Meynet03}
Meynet, G., \& Maeder, A. 2003, A\&A, 404, 975

\bibitem[{Meynet \& Maeder(2005)}]{Meynet05}
--- 2005, A\&A, 429, 581

\bibitem[{Oey(1996)}]{Oey96}
Oey, M.~S. 1996, ApJ, 465, 231

\bibitem[{Oey \& Clarke(2005)}]{OeyClarke05}
Oey, M.~S., \& Clarke, C.~J. 2005, ApJ, 620, 43

\bibitem[{Oey \& Massey(1995)}]{OeyLH47}
Oey, M.~S., \& Massey, P. 1995, ApJ, 452, 210

\bibitem[{Owocki \& van Marle(2008)}]{StanTheMan}
Owocki, S., \& van Marle, A.~J. 2008, in Massive Stars as Cosmic Engines,
  edited by F.~Bresolin, P.~A. Crowther, \& J.~Puls (Cambridge: Cambridge
  University Press), vol. 250 of IAU Symposium, 71

\bibitem[{Parker et~al.(1992)Parker, Garmany, Massey, \& Walborn}]{ParkerLH9}
Parker, J.~W., Garmany, C.~D., Massey, P., \& Walborn, N.~R. 1992, AJ, 103,
  1205

\bibitem[{Schaerer et~al.(1993)Schaerer, Meynet, Maeder, \&
  Schaller}]{Schaerer93}
Schaerer, D., Meynet, G., Maeder, A., \& Schaller, G. 1993, A\&AS, 98, 523

\bibitem[{Schaller et~al.(1992)Schaller, Schaerer, Meynet, \&
  Maeder}]{Schaller92}
Schaller, G., Schaerer, D., Meynet, G., \& Maeder, A. 1992, A\&AS, 96, 269

\bibitem[{Slesnick et~al.(2002)Slesnick, Hillenbrand, \& Massey}]{Slesnick}
Slesnick, C.~L., Hillenbrand, L.~A., \& Massey, P. 2002, ApJ, 576, 880

\end{thebibliography}

\end{document}